\documentclass[preprint,showpacs,preprintnumbers,amsmath,amssymb]{revtex4}
\usepackage{graphicx}
\usepackage{dcolumn}

\begin{document}

\preprint{APS/123-QED}
\title{Collinear cluster tripartition as sequential
binary fission in the $^{235}$U(n$_{\rm th}$,f) reaction}
\author{R. B. Tashkhodjaev}
\email{rustam@jinr.ru}
\affiliation{Institute of Nuclear Physics,  Tashkent, Uzbekistan}
\author{A.K. Nasirov}
\altaffiliation{Institute of Nuclear Physics,  Tashkent, Uzbekistan}
\email{nasirov@jinr.ru}
\affiliation{Joint Institute for Nuclear Research, Dubna, Russia}
\author{W. Scheid}
\affiliation{Institut f\"ur Theoretische Physik der
Justus-Liebig-Universit\"at, Giessen, Germany}
\date{Received: date / Revised version: date}

\begin{abstract}
The mechanism leading to the formation of the observed
products of the collinear cluster tripartition is carried out
within the framework of the model based on the dinuclear system
concept. The yield of fission products is calculated using the
statistical model based on the driving potentials for the
fissionable system. The minima of potential energy of the decaying
system correspond to the charge numbers of the products which are
produced with large probabilities in the sequential fission (partial
case of the collinear cluster tripartition) of the compound nucleus.
The realization of this mechanism supposes the asymmetric fission
channel as the first stage of sequential mechanism. It is shown that
only the use of the driving potential calculated by the binding energies
with the shell correction allows us to explain the yield of
the true ternary fission products. The theoretical model is applied
to research collinear cluster tripartition in the reaction $^{235}$U(n$_{\rm th}$,f).
Calculations showed that in the first stage of this fission
reaction, the isotopes  $^{82}$Ge and $^{154}$Nd are formed with
relatively large probabilities and in the second stage of sequential
fission of the isotope Nd mainly Ni and Ge are formed. This is in
agreement with the yield of the isotope $^{68}$Ni which is observed
as the product of the collinear cluster tripartition in the experiment.
\end{abstract}
\maketitle
\section{Introduction}

The observation of two and more nuclear fission products in the
fission of  $^{235}$U with thermal neutrons and in the spontaneous
fission of $^{252}$Cf has opened a new area of study in the
nuclear reactions. This phenomenon is connected with the appearance
of cluster states in nuclear reactions and it is the manifestation of
the shell structure which is responsible for the production of
isotopes with magic numbers of neutrons and protons. When a massive
nucleus loses its stability and goes to fission, first of all
clusters are formed as future fragments having the neutron or/and
proton number nearby the magic numbers 28, 50, 82 and 126. In the
case of ternary fission one observes fragments with the charge
number 28.

This work is devoted to the study of spontaneous and induced fission
of heavy nuclei. There is no full understanding of the fission
process and of the dependence of the probability of formation of
reaction products on the fission stages. Though one event of the
collinear cluster tripartition (hereinafter CCT) process occurs
against 1000 events of binary fission, the knowledge of its
mechanism gives us a better understanding of the process of
spontaneous and induced fission. A theoretical and experimental
knowledge of the nature of multicluster fission of various nuclei
will promote the construction of a full picture of fission.

The role of the nuclear shell structure in the formation of fission
products appears in the observed asymmetric mass distributions
depending on the full number of neutrons and the excitation energy
of the system undergoing to fission. The other manifestation of the
nuclear shell structure occurs at CCT
which was investigated by the FOBOS collaboration
\cite{Pyatkov2003,PyatkovEPJA45,PyatkovPHAN73,Pyatkov2011}. In this
paper, the authors attempt to explain reasons for the yield of the
observed fragments in CCT in the
$^{235}$U(n$_{\rm th}$,f) reaction. The cross sections of
multicluster fission of the U, Pu and Cf isotopes are less than one
percent of the corresponding cross sections of binary fission. So
the cross section of CCT is comparable with one of the well-known
ternary fission with the emission of an alpha particle. Therefore,
theoretical interpretation of these processes is required for a full
understanding of the mechanism.
The ternary fission fragmentation of $^{252}$Cf for all possible
third fragments using the recently proposed three-cluster
model \cite{ManimaPRC} was studied in Ref.\cite{ManimaEPJA}.
The authors concluded that the theoretical relative yields imply
the emission of the $^{14}$C, $^{34,36,38}$Si,
$^{46,48}$Ar, and $^{48,50}$Ca as the most probably third
particle in the spontaneous ternary fission of $^{252}$Cf.

\section{Theoretical model}

For the description of the formation of mass and charge
distributions of fission products of heavy nuclei the concept of the
dinuclear system (DNS) \cite{Antonenko1993,Volkov1999,Nasirov2005}
has been applied.
The formation of the DNS is an inevitable stage in fission
as in fusion. Indeed, at the descent from the saddle point to the
scission point, the fissile nucleus looks like a dumbbell and
changes its shape and mass (charge) asymmetry by nucleon exchange on
the way to the scission point.
The initial stage of fission when the shape of the
fissionable nucleus transforms from the compact form into the double
nuclear system is not considered here.
The charge distribution of the fragments is calculated
 after the crossing of the saddle point and  the formation of DNS.
In the case of the DNS with a large neck parameter the DNS model
is not acceptable because the DNS concept proposes the smallness of the
neck connecting the two interacting nuclei relatively to its whole
volume.

The similar conception was used in Ref.\cite{Andreev2006} to estimate
the yield of the $^4$He,  $^8$Be, and  $^{10}$Be in the ternary fission
of $^{252}$Cf and $^{56}$Ni. The authors of Ref. \cite{Andreev2006}
assumed the ternary fission as a two-step process:
the binary system is firstly formed and then the ternary system
is formed from it by extracting the light charged particle
($^4$He or  $^{8,10}$Be) into the region between the two heavy fragments.
Then this ternary system decays. According to their concept
the ternary system exists together for the short time then
it decays while in this work the heavy products undergoes to
fission after scission from the light fragment of the previous
binary fission of compound nucleus.

\begin{figure}
\par
\begin{center}
\resizebox{0.4\textwidth}{!}{\includegraphics{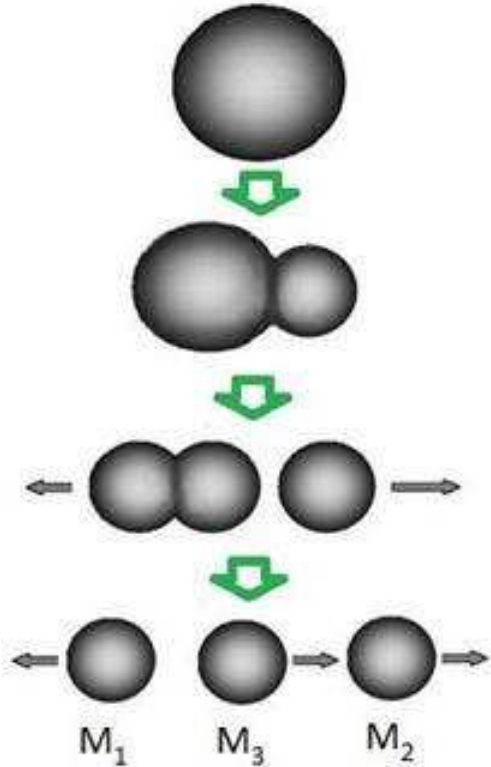}}
\end{center}
\vspace*{-1.0cm} \caption{The presentation of the sequential
collinear cluster tripartition of a heavy nucleus.} \label{graph1}
\end{figure}

CCT can be assumed to be the
two-stage fission of the sequential binary fissions. In Fig.
\ref{graph1}, a sketch of the sequential mechanism is presented.
It is obvious, that such a way of division is possible, when the initial
nuclei decay through the asymmetric channel.  It
should be stressed that the tripartition axes are collinear because
the fission axes of both stages are in coincidence: the heavy
fragment of the primary fission does not change the momentum
direction at its fission into two fragments forming the second and
third products of ternary fission. We do not exclude the fluctuation
of the fission axis. But this physical quantity is not explored by us.
Thus, in the consecutive ternary fission of the compound nucleus
first two fragments with asymmetric masses and charges are formed.
Then the heavy fragment decays again into two parts.

\begin{figure}
\vspace*{3.0cm}
\begin{center}
\resizebox{0.775\textwidth}{!}{\includegraphics{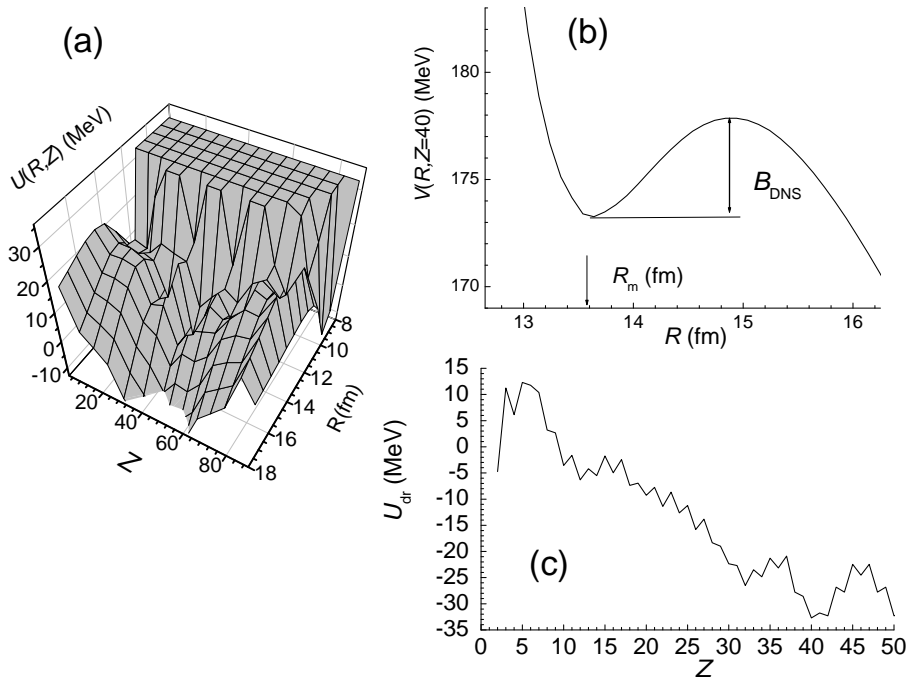}}
\end{center}
\vspace*{-4.3cm} \caption{Potential energy surface for $^{236}$U as
a function of the relative distance $R$ and a fragment's charge
number $Z$ (a); nucleus-nucleus potential $V(R)$ corresponding to
interaction of DNS $^{102}$Zr+$^{134}$Te. The depth
of the potential well is the barrier $B_{\rm DNS}$ against to decay
of DNS (b); driving potential for $^{236}$U (c).
} \label{graph2}
\end{figure}

 The mass and charge distributions of fission products of heavy nuclei
can be calculated if the descent from the saddle point up to the
scission points continues long enough. During the evolution of the
DNS along the mass asymmetry it should not decay into fragments. The
DNS concept assumes an interaction between two nuclei by nucleon
exchange between them and retaining their shell properties. The
validity of the application of the DNS concept to describe deep
inelastic transfer and quasifission reactions is evident: we have a
DNS in the entrance channel. But its application to explain the
yield of products in the fusion-fission, fast fission and
spontaneous fission processes requires more thoughts why it is
reasonable. With the DNS concept a number of features of deep inelastic
transfer reactions, fusion of nucleus, fission, quasifission,
fusion-fission and fast fission
\cite{Nasirov2005,Adamian1994,Nasirov2009} can be well explained.
The main degrees of freedom of the DNS models are the charge $Z$
and mass $A$ asymmetries of the system and the relative
distance $R$ between the fragments centers (see Fig. \ref{graph2}).
The mass number $A$ of a fragment was found from the minimization
of the total potential energy of the fissioning system at the
given charge number $Z$.
The DNS model supposes the knowledge of the interaction potential between
deformed nuclei with different orientation angles of their axial symmetry
axes \cite{Nasirov2005}. Details of the method of calculation of
 interaction potential is presented in Appendix.
This potential as a function of $R$ has a
well which allows DNS nuclei to be in interaction by multinucleon
transfer before the decay into two fragments. The depth of
this potential well corresponds to the scission point barrier which
was discussed in Ref.\cite{Royer}. This is one of the reasons to
explain the fission process. So the delay of scission is connected
by the potential well with a definite depth ($B_{\rm DNS}$,
see Fig. \ref{graph2}b).

The delay of the fission process results from the analysis of
experimental data showing a competition between particle (neutrons,
protons or alpha-particles) emission and fission of fusion-fission
reactions. To explain this phenomenon large friction forces or
viscosity of nuclear matter have been assumed \cite{Hinde1992}. It
 means  that the descent from the
saddle point up to the scission point occurs by nucleon exchange
between already formed nuclei during a long time
($(35\pm15)\times10^{-21}$s) in competition with the neutron
emission.
During the descent time, mass
(charge) equilibrium in DNS can be reached \cite{Moretto1975} by
nucleon exchange which is affected by the driving potential (see
Fig. \ref{graph2}c). The information which is used in our model
about crossing of the saddle point is the excitation energy
generated at the descent from there.

The effects of the relative motion between the fragments (elongation)
 and mass asymmetry vibration on the fission mass asymmetry
 were studied  in Ref. \cite{MaruhnPRC13}. The authors concluded
 that the static potential energy surfaces are
quite sufficient to determine the gross features of
fission.
In Ref. \cite{GuptaPRL},
the charge distribution of the reaction  fragments
with the fixed mass numbers at the spontaneous fission of $^{236}$U
was calculated by solving the Schroedinger equation
with the potential energy of the charge and mass numbers.
The authors used the total potential energy
which was obtained from the single-particle levels $\epsilon_i$
of the asymmetric two-center shell model by renormalizing
the sum  $\sum_{i=1}^{A}\epsilon_i$  in the Strutinsky method
\cite{Strutinsky} to the liquid-drop model of Myers and Swiatecki
\cite{Myers} with a modified surface asymmetry constant
\cite{Johansson}. The calculated nuclear charge dispersion in the fission
of U was in the good agreement with the experimental data for the
 mass fragmentation chains $A_1=141$, $A_2=95$ and $A_1=142$, $A_2=94$.

The DNS model supposes the change of the DNS total energy as change
of a sum of the energy reaction balance ($Q_{\rm gg}$-value) and
nucleus-nucleus interaction potential $V(R)$ between its
constituents:
\begin{equation}
U(Z,A,R)=V(Z,A,R)+Q_{gg}(Z,A).
\label{Utot}
\end{equation}
The $Q_{gg}$ - value represents the change of the internal energy of
the system during the reaction: $Q_{gg}=B_1(Z,A)+B_2(Z_{\rm CN}-Z,A_{\rm
CN}-A)-B_{\rm CN}(Z_{\rm CN},A_{\rm CN})$, where $Z_{CN}=Z_1+Z_2$;
$B_1$ and $B_2$ ($Z_1$ and $Z_2$) are the binding energies (charge
numbers) of the DNS constituents.  In Fig. \ref{graph2}a the
potential energy surface $U(Z,R)$ is presented as a function of the
relative distance between the fragments and the charge number $Z$ of
a fragment. The nucleus-nucleus potential $V(R)$ shown in Fig.
\ref{graph2}b corresponds to the interaction between fragments
$^{A_1}Z_1$ and $^{A_2}Z_2$. The charge numbers corresponding to the
mass numbers $A_1$ and $A_2$ provide minimal values of the potential
energy surface.

The driving potential has been found by connecting the minima of the
potential well of the nucleus-nucleus interaction (see Fig.
\ref{graph2}b) calculated for each charge asymmetry of the DNS,
{\it i.e.} it is the curve lying on the bottom of a valley of the
potential energy surface  along the $Z$ axis $2<Z<90$.
 If we use $R_m(Z)$ to show the value of $R$
corresponding to the bottom of the potential well for the DNS with the
charge asymmetry $Z$, then the driving potential is defined by using
(\ref{Utot}) as
\begin{equation}
U_{\rm dr}(Z,A)=V(Z,A,R_m(Z))+Q_{gg}(Z,A).
\label{Udr}
\end{equation}

The driving potential $U_{dr}$  describing the charge distribution
of the DNS fragments formed in the fission of $^{236}$U is shown in
Fig. \ref{graph2}c.  Advantage of such method of calculation of
the driving potential for the fissionable nuclear system is the
possibility to use the experimental values of binding energy of
nuclei $B_1$ and $B_2$ \cite{Audi2003} that allows us for the
account shell effects.  The shell effects play the crucial role in the
formation of the fission products. The maxima of their neutron and
proton numbers are close to the magic numbers which are inherent to
the clusters. In our calculations, for the values of binding energy
of isotopes which have not been measured yet, we use those obtained
from the well known mass tables \cite{Moller1988}.

The  peculiarities of the potential energy surface for
DNS allows us to find the basic directions of the evolution and main
modes of the decay.

\section{Application of the concept of the DNS for the explanation of binary fission}

  The DNS evolution is determined by
the potential energy surface and especially by the driving
potential. The driving potential sets the form and position of the
maximum of the mass (charge) distribution, as well as the total
kinetic energy of the products of the DNS decay. Therefore, the
correct description of the experimental data or their interpretation
depend on the accuracy of the calculation of the potential energy
surface of the DNS \cite{Pashkevich2008}. Large
probabilities of the formation of fragments with magic numbers 20,
28, 40, 50, 82 and 126 of protons and neutrons (clusters) are
obtained by including the quantum shell effects into the
consideration. Formation and yields of the fission products with
given mass and charge numbers are defined by the landscape of the
potential energy surface. We take into account the shell effects
to the binding energy of the DNS fragments to explain features of
the yields of products in asymmetric fission, which are observed in
experiment. The simplest way to take into account shell effects is
the use of binding energy values of atomic nuclei from the well
known tables \cite{Audi2003}.

The decay of the DNS may be analyzed as statistical process \cite{Andreev2006},
{\it i.e.} statistical equilibrium in the charge (mass) distribution of
the system can be established in dependence on the given excitation
energy and height of the barrier hindering its decay into two
fragments. In Ref. \cite{Andreev2006}, the ternary system with a light
nucleus between two heavy fragments is assumed to appear
from the binary con¯guration near scission. The theoretical results
of the authors for the charge
distributions of the light charged particles which are emitted
in spontaneous ternary  fission of $^{252}$Cf and in
induced ternary fission of $^{56}$Ni are in a good agreement
with the available experimental data.

The charge (mass) distribution of the fission fragments - yield
$Y(Z)$  depends on the charge distribution of the DNS fragments
$P(Z)$ and decay probability $W(Z)$ of the DNS from the given charge
asymmetry state $Z$:
\begin{equation}
Y(Z)=Y_{0}P(Z)W(Z),
\label{Yz}
\end{equation}
where $Y_{0}$ is a normalizing coefficient for the yield
probabilities. The probability of formation of the DNS $P(Z)$ can be
found from the condition of a statistical equilibrium as in Ref.
\cite{Moretto1975}:
\begin{equation}
P(Z)=P_{0}e^{-U_{\rm dr}(Z)/T_{\rm DNS}(Z)}
\label{Pz}
\end{equation}
where  $T_{\rm DNS}(Z)$ is the effective temperature of the DNS with
the charge asymmetry $Z$ and $U_{dr}(Z)$ is determined by
formula (\ref{Udr}).  The
decay probability of the DNS $W(Z)$ can be found as in Ref.
\cite{Andreev2006}:
\begin{equation}
W(Z)=W_{0}e^{-B_{\rm DNS}(Z)/T_{B}(Z)},
\label{Wz}
\end{equation}
where $T_{B}(Z)$ is the effective temperature of the system on the
barrier (or at the scission point), $B_{\rm DNS}(Z)$ is the
scission barrier for the decay of the DNS
(see Fig. \ref{graph2}b). Its value is determined by the depth
of the potential well of the nucleus-nucleus interaction.

From the equations (\ref{Yz})-(\ref{Wz}) we can see that the yield
of the fragments at the decay of the DNS depends strongly on the
driving potential $U_{dr}(Z)$. The equation (\ref{Pz}) means that
the position of the maxima of the mass distribution corresponds to
the minima of driving potential, which is calculated for a given
decaying nucleus. Therefore,  we consider the minima of the driving
potential. For the theoretical research of the yield of fission
fragments, which are observed in experiment, the driving potential
must be defined sufficiently accurately.

\section{Explanation of ternary fission as cascade fission in
the $^{235}$U(n$_{\rm th}$,f) reaction}

The knowledge about the driving potential $U_{\rm dr}$ of a
fissionable system allows us to make predictions concerning the
shape of the mass distribution of fission products. There is the
possibility of the explanation of the ternary fission process with
specialities of the driving potential. We have considered the
reaction $^{235}$U(n$_{\rm th}$,f) in which CCT was studied in Ref.
\cite{PyatkovEPJA45,PyatkovPHAN73}. According to the sequential
mechanism the fission of the heaviest fragment in the first fission
stage produces the second and third fragments. Firstly, we shall
consider the first binary fission
\begin{center}
n$_{\rm th}$+$^{235}$U$\rightarrow$$^{236}$U$^{*}$$\rightarrow$
F$_{1}$+F$_{2}$
\end{center}
by the thermal neutron with energy $E_{\rm n_{th}}=0.025$ eV. The
excitation energy of the compound nucleus $^{236}$U$^{*}$ is 6.54
MeV. The potential energy surface is calculated by formula
(\ref{Utot}) which includes the shell effects due to the use of
realistic binding energies of the interacting nuclei.  The energy
balances for the different fission modes (F$_{1}$+F$_{2}$) are
presented in Table 1. According to the total energy conservation law
the total kinetic energy should not be larger than the energy balance
($Q_{\rm gg}$ - value) in the corresponding channel. Therefore, in
the calculation of the potential energy surface for the
$^{236}$U$^{*}$$\rightarrow$ F$_{1}$+F$_{2}$ reaction the static
deformation parameters of the constituents of DNS were used as free parameters
to make the barrier of the nucleus-nucleus potential for the exit
channel lower than the maximal value of $TKE$.
The driving potential has been
calculated with the formula (\ref{Udr}) and the result is presented
in Table 1. The theoretical value of  $TKE$ included the energy
carried away by 2 neutrons which are emitted. Therefore, there is a
difference between theoretical and experimental
values of  $TKE$ for the $^{92}$Kr+$^{144}$Ba fission channel.\\

\noindent
\small{Table 1.  Energy balance ($Q_{\rm gg}$) and theoretical
($TKE(\rm theor)$) and experimental ($TKE(\rm exp)$) values \cite{Pyatkov2002}
of the total
kinetic energy (in MeV) of the  fragments for the different
fission modes of $^{236}$U (the first stage). $B^H_f$ is the fission
barrier of the heavy fragment formed in the first stage of the
sequential fission. $\beta_2^{(1)}$ and $\beta_2^{(2)}$ are the
quadrupole deformation parameters of the first and second
fragments, respectively, which correspond to the $2^+$-state and are
taken from Ref. \cite{Raman1987}.}

\begin{center}
\begin{tabular}{|l|c|c|c|}
\hline
 Fission modes & \small  $^{82}$Ge+$^{154}$Nd & \small
 $^{86}$Se+$^{150}$Ce & \small $^{92}$Kr+$^{144}$Ba\\
\hline
$Q_{\rm gg}$(MeV) & -173.75 & -177.976 & -193.912\\
\hline
$TKE(\rm theor)$ & 159.83 & 168.46 & 190.69\\
\hline
$TKE(\rm exp)$ &  &  & 170.0 \cite{Pyatkov2002} \\
\hline
$B^H_f$(MeV) & 36.33 & 37.92 & 39.17 \\
\hline
$\beta_2^{(1)}$ & 0.26 & 0.19 & 0.15 \\
$\beta_2^{(2)}$ & 0.35 & 0.32 & 0.19 \\
\hline
\end{tabular}
\end{center}

\begin{figure}
\par
\begin{center}
\resizebox{0.85\textwidth}{!}{\includegraphics{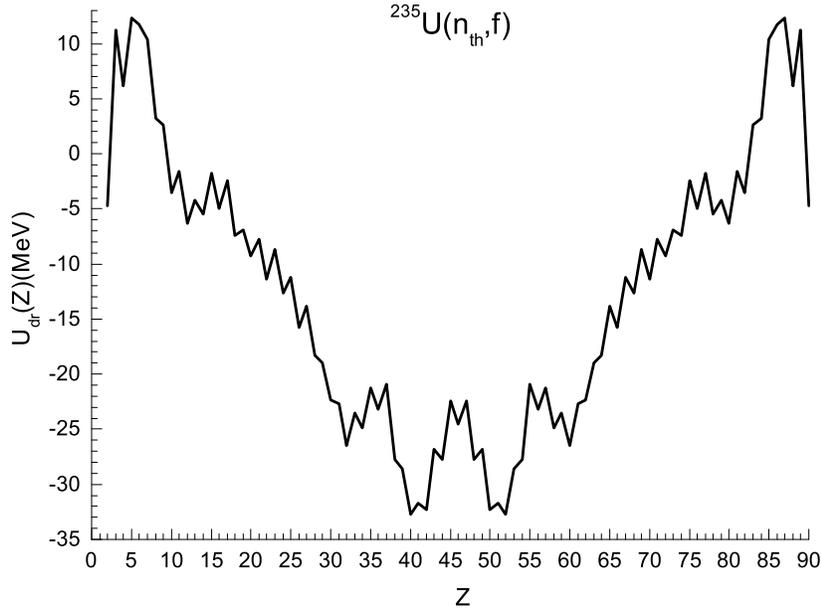}}
\end{center}
\vspace*{-1.6cm} \caption{Driving potential for the
$^{235}$U(n$_{\rm th}$,f) reaction calculated by the use of the
binding energies with the shell effects.} \label{graph3}
\end{figure}
\begin{figure}
\par
\begin{center}
\resizebox{0.85\textwidth}{!}{\includegraphics{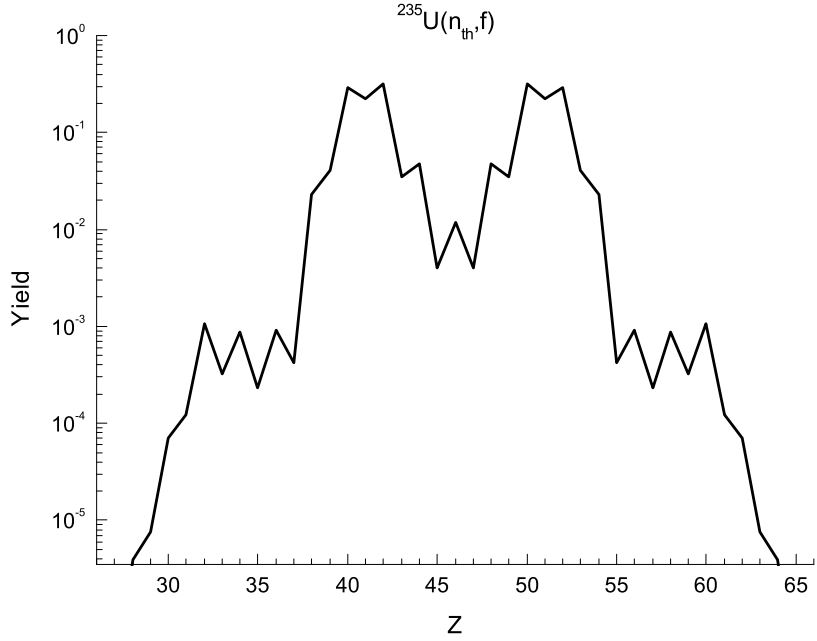}}
\end{center}
\vspace*{-1.6cm} \caption{Yields of the reaction products at fission
of $^{236}$U calculated with formula (\ref{Yz})} \label{graph4}
\end{figure}
\begin{figure}
\par
\begin{center}
\resizebox{0.85\textwidth}{!}{\includegraphics{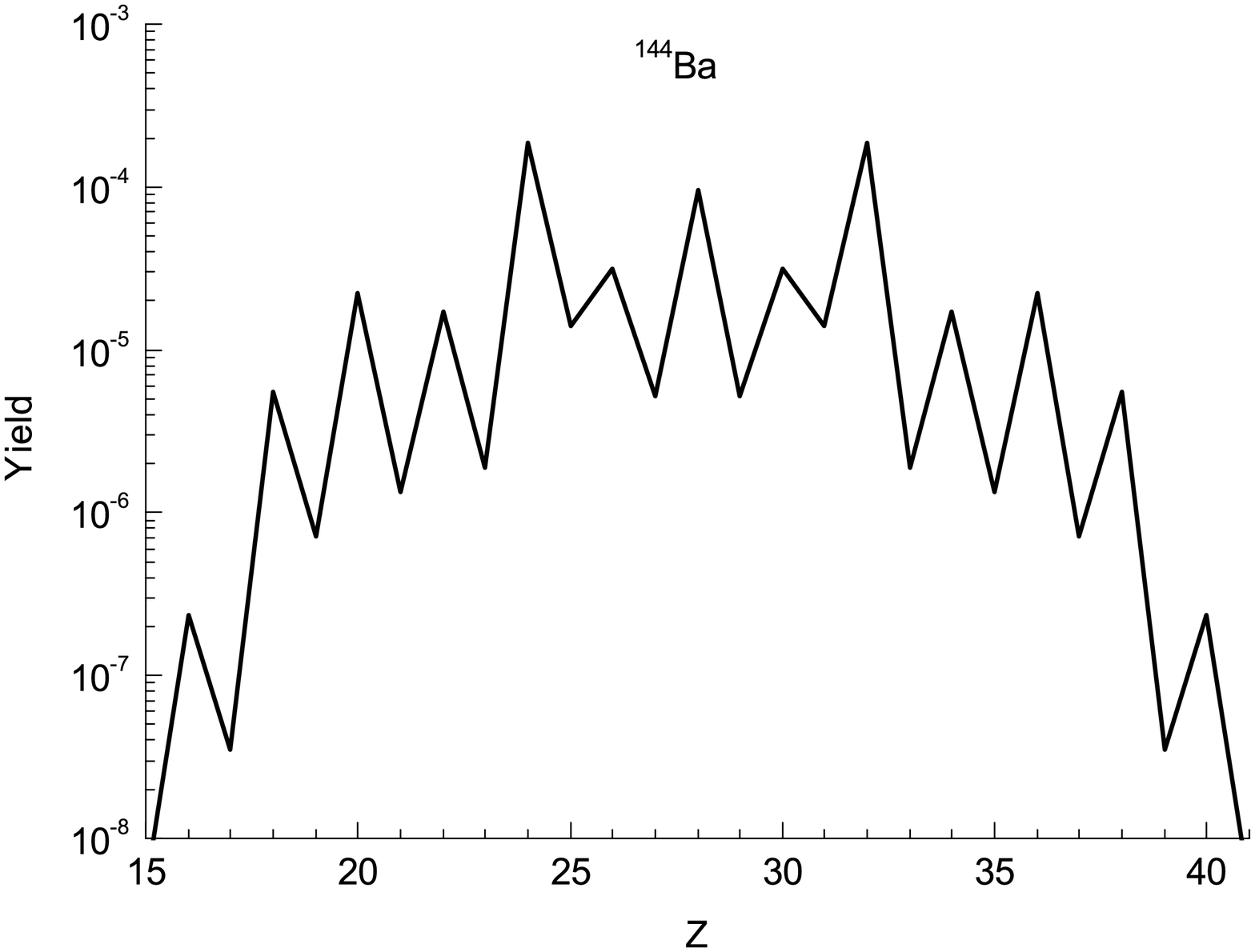}}
\end{center}
\vspace*{-1.6cm} \caption{Yields of the reaction fragments for
fission of $^{144}$Ba.} \label{graph5}
\end{figure}
\begin{figure}
\par
\begin{center}
\resizebox{0.85\textwidth}{!}{\includegraphics{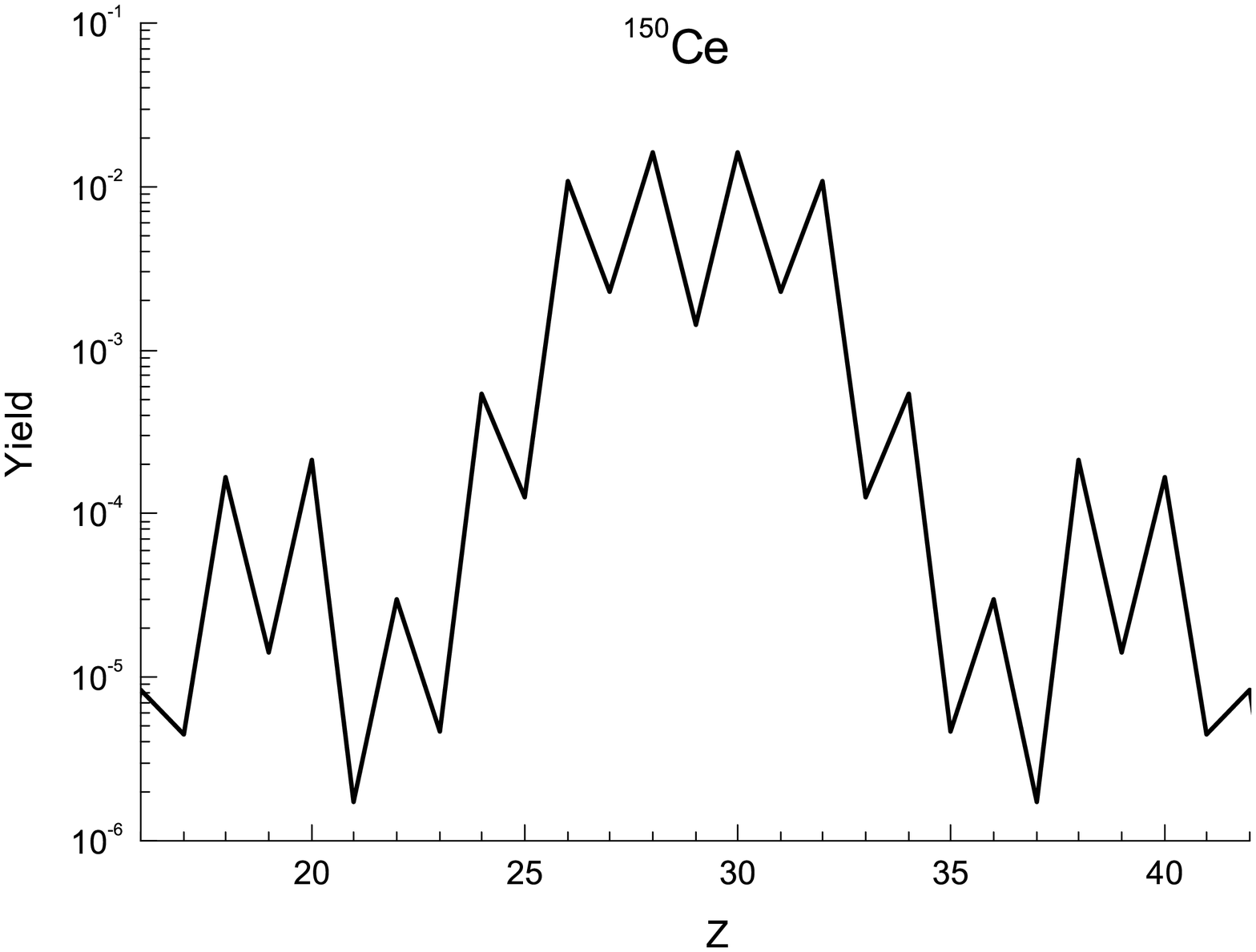}}
\end{center}
\vspace*{-1.6cm} \caption{Yields of the reaction fragments at
fission for the $^{150}$Ce.} \label{graph6}
\end{figure}
\begin{figure}
\par
\begin{center}
\resizebox{0.85\textwidth}{!}{\includegraphics{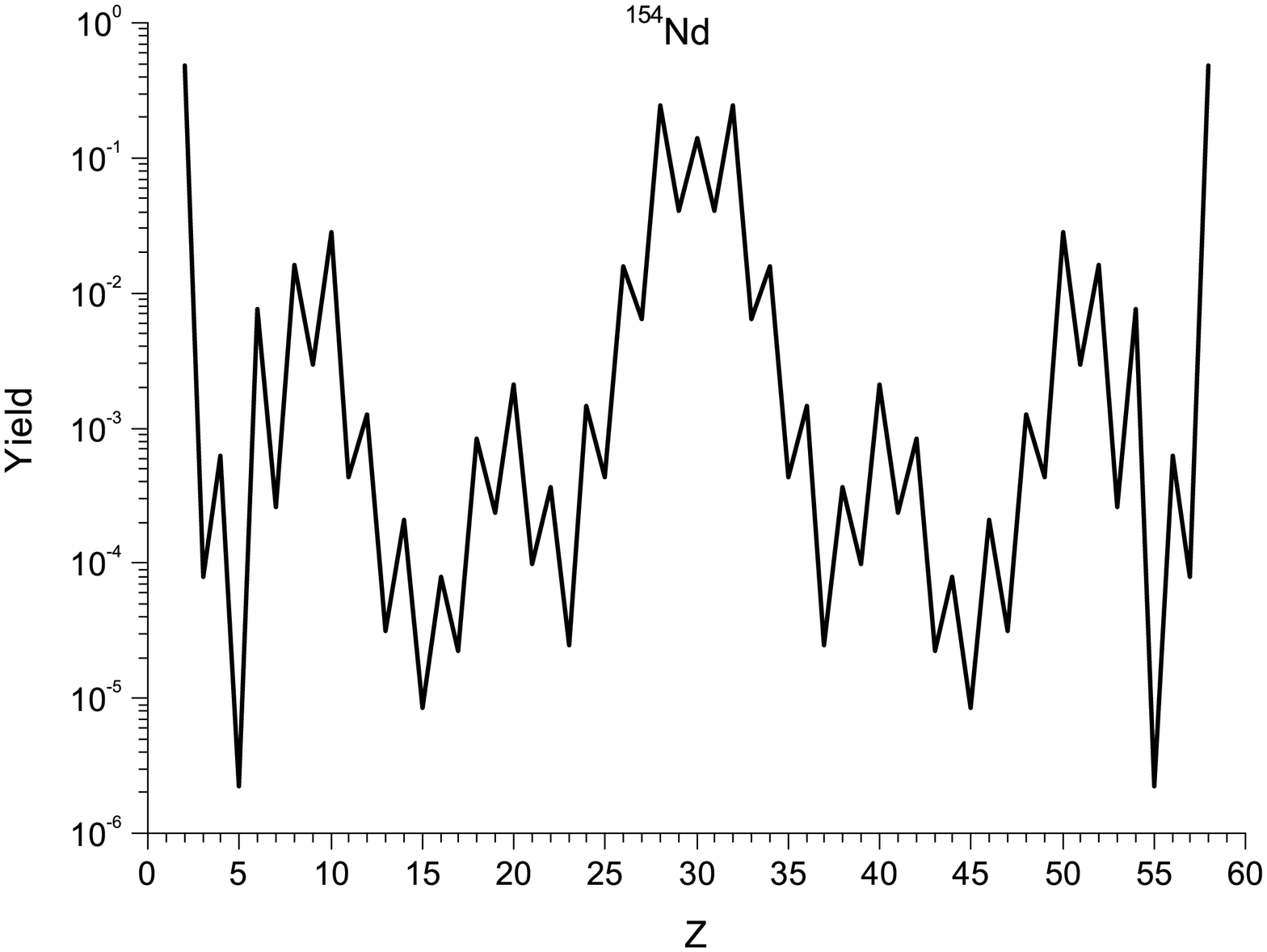}}
\end{center}
\vspace*{-1.6cm} \caption{Yields of the reaction fragments at
fission for the $^{154}$Nd.} \label{graph7}
\end{figure}

 As shown in Fig. \ref{graph4} the yields of the
products with the charge numbers $Z$=10, 28, 30, 32, 50, 56, 58 and
60 are well pronounced. In Fig. \ref{graph3} three minima located at
$Z$=56, 58, 60 are responsible for the corresponding maxima
at these charge numbers in Fig. \ref{graph4}. In the main fission channels light fragments
with the most probable charge numbers $Z$=40 and 42 and the
corresponding heavy fragments with $Z$=52 and 54 are produced with larger
fission barriers in comparison with the one of the heavier fragments
$Z$=58 and 60 (see Table 1). The barrier values are obtained for the
ground state of nuclei by the Sierk's model with the rotating
liquid-drop model \cite{Sierk1986}. The shape of fission products is
not expected to be in the ground state: as soon as they are prolate
deformed, the fission barriers are smaller than presented in Table
1.  So, we will consider the fragments Ba, Ce, Nd as fissionable
nuclei in the second stage of the CCT.
\[\textrm{n}_{\rm th}+^{235}\textrm{U} \rightarrow ^{236}\textrm{U}^{*}
\rightarrow \left\{ \begin{array}{ll} ^{92}\textrm{Kr}+^{144}\textrm{Ba}\\
 ^{86}\textrm{Se}+^{150}\textrm{Ce}\\
 ^{82}\textrm{Ge}+^{154}\textrm{Nd}
 \end{array}
 \right.
 \]
The results of the calculations for the yields of fragments in the
fission of Ba, Ce and Nd are presented in Figs \ref{graph5},
\ref{graph6} and \ref{graph7}, respectively.
\begin{figure}
\par
\begin{center}
\resizebox{0.85\textwidth}{!}{\includegraphics{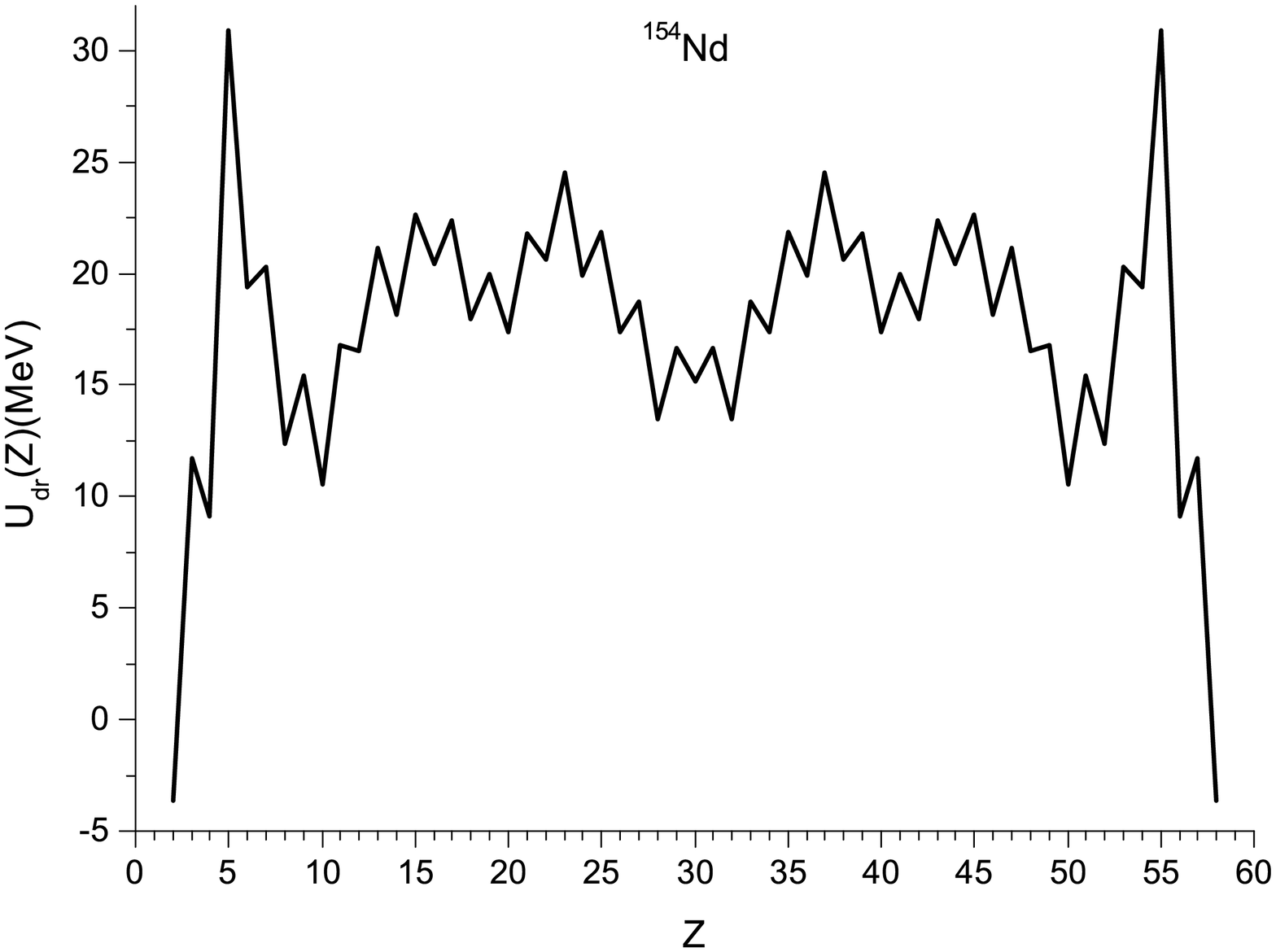}}
\end{center}
\vspace*{-1.6cm} \caption{Driving potential used to calculate the
yields of the fission fragments of $^{154}$Nd.} \label{graph8}
\end{figure}

The comparison of the yields of the fission of Ba, Ce, Nd shows that
the maxima of the probability of the yields corresponding to the
products in CCT
$Y(^{144}$Ba)=$10^{-4}\div 10^{-3}$ in the fission of Ba,
$Y(^{150}$Ce)=$10^{-2}\div 10^{-1}$ in the fission of Ce and 
$Y(^{154}$Nd)=$0.1\div 1$ in the fission of Nd. It means that the
formation of the intermediate nucleus Nd and its sequential fission
can be considered as the second stage of the ternary sequential
fission to interpret the observed yield of the Ge and Ni isotopes
\cite{PyatkovEPJA45,PyatkovPHAN73}. The neutron emission from the
fission fragments was calculated as in Ref. \cite{Andreev2007}:
\begin{equation}
\nu_i=\frac{E^*_i}{B^{(n)}_i+2T_i}
\end{equation}
where $B^{(n)}_i$ is the separation energy of the neutron in the
fragment $i$ and $T_i$ is the temperature of the fragment. In Figs.
\ref{graph7} and \ref{graph8}, the yields of reaction products for
the fission of $^{154}$Nd and the corresponding driving potential
are presented. The results of the calculations show that $^{82}$Ge and
$^{154}$Nd can emit one and two neutrons, respectively. The fission
of $^{154}$Nd competes with the emission of two neutrons.
The fission probability after
neutron emission is smaller then the one before neutron emission
because the excitation energy of the fissioning nucleus decreases by
neutron emission. In the second stage of ternary fission the nucleus
$^{154}$Nd fissions into two fragments $^{72}$Ni and $^{82}$Ge.
The maxima of the charge and mass distribution of the
second stage fission products lie at these reaction products.
Two neutrons are emitted from the $^{72}$Ni nucleus: the two
neutrons separation energy is equal to 10.93 MeV. The similar energy
for the two neutron emission from the $^{82}$Ge nucleus is larger, namely
12.25 MeV because $^{82}$Ge is a double magic nucleus. Thereby, we
conclude the following mechanism of CCT of the excited $^{236}$U:
at the first stage it
decays into the $^{82}$Ge and $^{154}$Nd products. The
excited nucleus Ge emits 1 neutron. In the second stage the $^{154}$Nd
nucleus decays into two fragments  $^{72}$Ni and $^{82}$Ge,
and the excited $^{72}$Ni emits 2 neutrons.
The charge symmetric fragmentation
$^{154}$Nd$\rightarrow^{76}$Zn+$^{78}$Zn channel has a smaller
probability than the $^{72}$Ni+$^{82}$Ge channel. Therefore, our
calculations show that the main channel of CCT of the $^{236}$U$^{*}$ is
$^{81}$Ge+$^{70}$Ni+$^{82}$Ge+3n. The probability of the ternary fission into
this channel is
$1.06\cdot10^{-3}\times2.46\cdot10^{-1}\approx3\cdot10^{-4}$.
The relative probability of CCT to binary fission is approximately
$10^{-3}$.
\begin{figure}
\par
\begin{center}
\resizebox{0.85\textwidth}{!}{\includegraphics{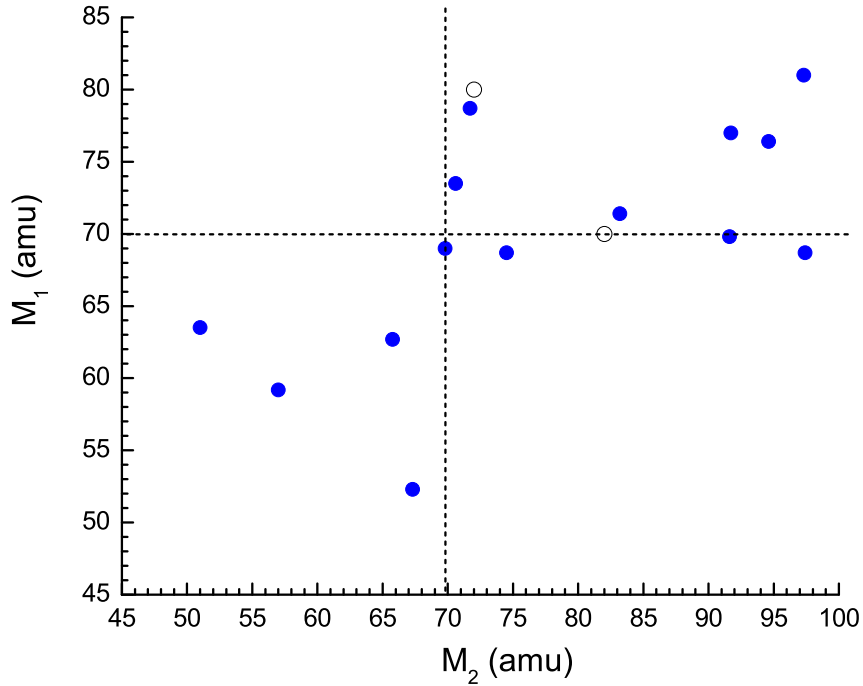}}
\end{center}
\vspace*{-1.6cm} \caption{
Comparison of the maximum values of the yield of the tripartition products
(open circles) corresponding to the masses \{$M_1=80$, $M_2=72$\} and
\{$M_1=70$, $M_2=82$\} which indicate the yield of $^{70,72}$Ni
and $^{80,82}$Ge with the experimental data of the mass-mass distribution
(filled circles, from Ref. \cite{PyatkovPHAN73}) of the $^{236}$U(n$_{\rm th}$,f)
fission fragments registered in coincidence.} \label{graph9}
\end{figure}

In Fig. \ref{graph9}, the maximum values of the mass distributions
of the two pair CCT products \{$M_1=80$, $M_2=72$\} and
\{$M_1=70$, $M_2=82$\} obtained in this work
for the $^{236}$U(n$_{\rm th}$,f) reaction are compared
 with the experimental data of the mass numbers of the two
products of tripartition taken from Fig. 6b of
Ref. \cite{PyatkovPHAN73}. Authors of Ref. \cite{PyatkovPHAN73}
used the  ``lost fragment'' method, when the setup can register
only two fragments and the third fragment of tripartition is missed.
The experimental data present the
mass–mass distribution of the fission fragments
corresponding to the selected events with approximately equal
momentum and velocity distributions, as well as with similar
nuclear charges.  It is seen from
this figure that the theoretical yields of the $^{70,72}$Ni isotopes accompanied
with the yields of the $^{81,82}$Ge isotopes are close to the
experimental data presented in  Ref. \cite{PyatkovPHAN73}. It means that
these experimental events correspond to the sequential two stage mechanism of
CCT which is shown in Fig. \ref{graph1}. The analysis of the yields of products
having sufficient probabilities in other channels of the first stage fission
 $^{236}$U$^*\rightarrow^{86}$Se+$^{150}$Ce should be investigated and may add
new theoretical results.

\section{Total kinetic energy in fission}

During the whole process the energy conservation rule must be
fulfilled.  After the neutron capture we have an excited compound
nucleus with the total energy
\begin{equation}
E_{CN}^{*}+B_{CN}=B_1+B_2+E_{1}^{*}+E_{2}^{*}+V(Z,A,R)+K_1+K_2,
\end{equation}
where $E_i^*$  and $K_i$ ($i$=1,2) are the excitation energies and
kinetic energies of the constituents of the DNS.

The total potential energy surface is determined according Eq.
(\ref{Utot}). The total kinetic energy ($TKE$) and excitation energy of the dinuclear
system are determined by the sum of the corresponding energies of the
fragments $K_1+K_2=TKE$. The total kinetic energy  of the fragments
is restricted by the value which is found from the total energy
conservation law:
\begin{equation}
TKE(Z,A,R)=E^*_{CN}-U(Z,A,R)-E_{1}^{*}-E_{2}^{*}
\end{equation}
At large values of $R$ there is no interaction, {\it i.e.}
$V(R\rightarrow 0)$
\begin{eqnarray}
TKE(Z,A,R\rightarrow\infty)&=&E^*_{CN}-Q_{gg}(Z,A)\nonumber\\
&-&E_{1}^{*}(Z,A)-E_{2}^{*}(Z,A)
\end{eqnarray}
$E^*_{CN}$ is the excitation energy of the compound nucleus.  The value of
$TKE$, which includes kinetic energy of the relative motion and surface
vibrational energies of the nuclei, depends on the value of
$E_{1}^{*}+E_{2}^{*}$. One can say that the fixed intrinsic energy
$E^*_{CN}-Q_{gg}$ is distributed between $TKE$ and
$E_{1}^{*}+E_{2}^{*}$. The latter energy may be spent for emission
of nucleons from the system and for its deformation.

For the initial stage, when $^{236}$U is formed by the capture of a
neutron by $^{235}$U, the excitation energy of the compound nucleus is
determined from the energy balance of the reaction:
$E^*_{CN}(^{236}$U)=$E_{n_{th}}+Q_{gg}=E_{n_{th}}+B_1(n)+
B_2(^{235}$U)-$B_{CN}(^{236}$U)=6.5
MeV. As we discussed above, in the first fission stage of
$^{236}$U$^*$ two excited fragments are formed: $^{154}$Nd$^*$ and
$^{82}$Ge$^*$. The excitation energy of $^{154}$Nd$^*$ is calculated
from the assumption of a full thermodynamic equilibrium between the
formed two fragments
$E_{CN}^*(^{154}$Nd$)=\frac{A_{Nd}}{A_U}(U_{dr}(BG)-U_{dr}(Z=60))=25.3$
MeV, where $U_{dr}(BG)$ is the driving potential at the Businaro-Gallone point
of $^{236}$U which is equal to 12.3 MeV;
$U_{dr}(Z=60)=-26.5$ Mev is the  value of the driving potential for
$Z=60$  (see Fig. \ref{graph3}).  We did not calculate the fission
probability for $^{154}$Nd$^*$ but its excitation energy is enough
for fission of $^{154}$Nd$^*$. The charge and mass distributions of
its fission fragments are determined by the driving potential
presented in Fig. \ref{graph8}.

As shown in Table 1., the total kinetic energy of the fragments
at the first stage of CCT
$TKE(^{82}$Ge+$^{154}$Nd) is 159.83 MeV, and analogously we can
find the total kinetic energy for the second stage
$TKE(^{72}$Ni+$^{82}$Ge)=72.9 MeV.

The driving potential of the $^{235}$U(n$_{\rm th}$,f) reaction has
minima corresponding to magic numbers of the protons or neutrons equal to
2, 8, 20, 28, 50, 82, but the decay probability depends on the
splitting barrier height $B_{DNS}(Z)$ as a function of the charge
number. From this dependence it is clearly seen that the reaction
products Ni, Sn and Ge are clusters.

\section{Conclusions}

The sequential fission mechanism of ternary fission of a heavy
atomic nucleus has been considered within the framework of the DNS
model. Herewith, first the mother nucleus decays into two not alike
fragments (asymmetrical fission mode). Then the heavy product decays
into two further fragments. The axes of  both fission events are in
coincidence according to the momentum conservation rule if we assume
that the averaged initial angular momentum of the compound nucleus
$^{236}$U generated by thermal neutrons is very small. The purpose
of this work was an estimation of the values of charge and mass
numbers of the ternary fission  products  which are formed with large
probabilities and the comparison with the experimental
maxima in the charge and mass distributions.
For the estimation of these values we calculated the driving
potential and yields of fragments of fissionable nuclei. The
potential energy is introduced as the sum of the balance energy of the
reaction ($Q_{\rm gg}$-value) and the nuclear interaction potential of the
DNS constituents. The use of real binding energies of nuclei,
taken from Ref. \cite{Audi2003}  allows us directly to
take into account the shell (quantum) effects in atomic nuclei. Due
to the shell effects there is the possibility of the formation
of clusters having charge or neutron numbers near the magic numbers 28,
50, 82  in binary and ternary fission of nuclei.
The DNS allows  to take into account the shell effects in the
calculation of the potential energy surface of the fissioning system.
The given
method can be applied for the description of the yields of the products
of the ternary fission $^{236}$U in reactions $^{235}$U(n$_{\rm
th}$,f). As primary channel of the sequential
ternary fission we took the channel $^{82}$Ge+$^{154}$Nd.
After scission the nucleus $^{82}$Ge can evaporate one neutron
and it can be registered as the first fission fragment.
At the same time, the heavy fragment $^{154}$Nd is assumed to
undergo to fission into the channel $^{154}$Nd$\rightarrow^{72}$Ni+$^{82}$Ge.
 The neutron emission from the primary fragments $^{72}$Ni and $^{82}$Ge leads to
the formation of the reaction products $^{70}$Ni and $^{82}$Ge. This
result explains the observed yields of the ternary fission products Ni
and Ge with a relatively large cross section.
 Thereby, these estimations within the framework
of the DNS model indicate enough well the main channels of CCT.
 This means, that the chosen theoretical model can
be extended for the prediction of quantitative results for the mass
distributions of the products of fission.

\textbf{Acknowledgements}

The authors are grateful to Drs. D. V. Kamanin and Yu. V. Pyatkov
for valuable discussions. The authors thank DAAD and RFBR for the
partial support and Institute of Theoretical Physics of the
Justus-Liebig-University Giessen for the warm hospitality.

\begin{center}
\textbf{Appendix}
\end{center}

This method of calculation is applicable at the stage of fission
when two fragments are connected with a neck of a small size in
comparison to the whole size of the DNS. In this case the interaction
between the fragments of the DNS may be described by the diabatic
nucleus-nucleus potential for well deformed nuclei with a
repulsive core and gives the relaxation of the charge
asymmetry degree of freedom.

We use the nucleus-nucleus potential consisting of three parts:
\begin{equation}
V(R,Z,A,l,\beta)=V_{nuc}(R,A,\beta)+V_{C}(R,Z,\beta)+V_{rot}(R,Z,l),
\end{equation}
where $V_{nuc}$  and $V_C$ are the nuclear and Coulomb parts of the
nucleus-nucleus potential, respectively; $V_{rot}$ is the rotational
energy of the DNS. In ternary fission of $^{236}$U caused by thermal
neutrons discussed in this work the rotational energy can be
neglected due to the smallness of the partial wave number $l$.
Therefore, the rotation of the DNS is not considered in this work.

The nuclear part $V_{nuc}$ is calculated by the double folding potential:
\begin{equation}
V_{nuc}(R)=\int\rho_{1}(r')f_{\rm eff}[\rho(r,r')]\rho_{2}(r)d\textbf{r},
\label{Vnuc}
\end{equation}
\begin{center}
\textbf{r}$'$=\textbf{r}-\textbf{R},
\end{center}
where $\rho_1$ and  $\rho_2$ are the nucleon density distributions of
the interacting nuclei; $f_{eff}$ is the effective nucleon-nucleon
potential taken from Ref. \cite{Migdalbook}. The advantage of this
Migdal forces is their dependence on the nuclear density of the nuclei:
\begin{equation}
f_{\rm eff}[\rho(r,r')]=C\left[f_{\rm in}+(f_{\rm ex}-f_{\rm in})
\frac{\rho_{0}-\rho(r,r')}{\rho_{0}}\right],
\end{equation}
where $C$=300 MeV$\cdot$fm$^{3}$, $f_{in}$=$0.09$ and
$f_{\rm ex}$=$-2.59$ are constants from Ref. \cite{Migdalbook}. When
$\rho(r,r')>\rho_0$ the nuclear part becomes repulsive that
corresponds to the appearance of the Pauli blocking principle. For the
nucleon density distribution of nuclei we use Fermi functions
placed at the center-of-mass of the nuclei which have the distance $R$
between them,
\begin{equation}
\rho_{1}(r)=\frac{\rho_{0}}{1+\exp\left(\frac{r-R_{1}}{a}\right)},
\end{equation}
\begin{equation}
\rho_{2}(|\textbf{r}-\textbf{R}|)=\frac{\rho_{0}}{1+
\exp\left(\frac{|\textbf{r}-\textbf{R}|-R_{2}}{a}\right)}
\end{equation}
and
\begin{equation}
\rho(r,|\textbf{r}-\textbf{R}|)=\rho_{1}(r)+\rho_{2}(|\textbf{r}-\textbf{R}|).
\end{equation}
Here, $R_{i}$=$r_{0}A_{i}^{1/3}(1+\beta_{2}^{(i)}Y_{20})$ is the
radius of the $i^{th}$ nucleus, $r_0=1.15\div1.18$ fm, $\rho_0=0.17$
fm$^{-3}$, $a=0.54$ fm, and $\beta_{2}^{(i)}$ are the quadrupole
deformation parameters. Orientation angles of
the axial symmetry axes of the interacting nuclei are taken as $\Theta_1=0^{\circ}$
and $\Theta_2=180^{\circ}$ in the both stages of the sequential binary fission.

The integral in Eq. (\ref{Vnuc}) is calculated with numerical
methods.

The Coulomb potential between the nuclei of the DNS is found by the
formula of Wong \cite{Wong1973}:

\begin{eqnarray}
V_{C}(R,Z_{1},Z_{2})&=&\frac{Z_{1}Z_{2}e^{2}}{R}+\frac{Z_{1}Z_{2}e^{2}}{R^{3}}\sqrt{\frac{9}{20\pi}}\sum_{i=1}^{2}R_{0i}^{2}\beta_{2}^{(i)}\nonumber\\
&+&\frac{Z_{1}Z_{2}e^{2}}{R^{3}}\frac{3}{7\pi}\sum_{i=1}^{2}R_{0i}^{2}(\beta_{2}^{(i)})^{2}.
\end{eqnarray}
Here, $R_{0i}$ is the radius of the $i^{th}$ spherical nucleus.

\end{document}